\documentclass[final,3p,times,twocolumn]{elsarticle}

\usepackage{lineno,hyperref}
\usepackage{amssymb}
\usepackage{pifont}
\usepackage{amsmath}
\usepackage{textgreek}
\usepackage{float}
\usepackage{cleveref}
\usepackage{subcaption}
\usepackage{ifluatex}
\usepackage{graphicx} 
\usepackage{multirow}
\usepackage{textcomp}

\ifluatex
\usepackage{pdftexcmds}
\makeatletter
\let\pdfstrcmp\pdf@strcmp
\let\pdffilemoddate\pdf@filemoddate
\makeatother
\fi
\usepackage{svg}
\modulolinenumbers[5]

\begin{document}
	
	\begin{frontmatter}
		
		\title{Hydrogen Storage Performance Enhancement and Bandgap Opening of M-Decorated ($\mathrm{M}=\mathrm{Li}$, Na and K) III\textsubscript{4}-V\textsubscript{4} Monolayer by Fluorine Functionalization}
		
		\author[1]{Alireza Kokabi}
		\author[1]{Shoeib Babaee Touski\footnote[1]{Corresponding author: touski@hut.ac.ir}}
		\address[1]{Department of Electrical Engineering, Hamedan University of Technology, Hamedan 65155, Iran}
		
		

\begin{abstract}
The effects of fluorine functionalization on the hydrogen storage capability of alkaline decorated III\textsubscript{4}-V\textsubscript{4} monolayers is studied.
This structure can store up to two hydrogen molecules per alkaline atom. Here, we demonstrate that functionalizing alkaline decorated monolayer with a high electronegative element of fluorine can significantly enhance both binding energy and the maximum number of the stored hydrogen molecules. In this regard, the hydrogen molecule storage capability is notably improved from two to four and when absolute binding is considered. In addition, F-functionalization of M-decorated III\textsubscript{4}-V\textsubscript{4} can demonstrate bandgap opening effect, introduce semiconducting characteristics and forming a new two-dimensional semiconductor structure. The bandgap for Li-Al$_4$P$_4$-F is 1.23 eV which is very close to the solar peak. The resulted bandgap in the M-III\textsubscript{4}-V\textsubscript{4}-F structure is even significantly larger than that of pristine III\textsubscript{4}-V\textsubscript{4} monolayer. The binding of the hydrogen molecule to the alkaline atom is also improved from 0.114 to 0.272 eV by the fluorine functionalization.

\end{abstract}
	


\end{frontmatter}

\section{Introduction}

In recent years, the two-dimensional materials have attracted lots of
attentions due to their unique layered structures, large specific surface
areas and outstanding physical/chemical properties \cite{fiori2014electronics,xu2013graphene}. They have been successfully applied in the fields of supercapacitor electrodes \cite{liu2017recent,haider2020structural},
lithium/sodium-ion capacitors \cite{han2019two}, biomedical engineering \cite{chimene2015two,huang2021two}, chemical sensors \cite{varghese2015two} and energy storages \cite{zhang2018two,xue2017opening,peng2016two}.

The two-dimensional materials were also reported to be possible substrates to adsorb and store the gas molecules \cite{rao2015comparative,chettri2021hexagonal}. Among the different gas storages, hydrogen one has gained a special attraction for application as a clean fuel sources \cite{hu2014two}. Hydrogen can be consumed in the fuel cells and the resulting material would be water. 

Monolayers are promising candidates for hydrogen storage applications due to their high surface-to-volume ratio. However, most 2D materials demonstrate a small amount of binding to hydrogen molecules. 
Metal adatoms are suggested as a potent method to strengthen the hydrogen storage capability of 2D materials \cite{zhang2016ti,wang2018first,chen2017decorated,wei2021hydrogen}. Transition metals, alkali and alkaline-earth metals are particularly interested in this regard \cite{arellano2021ab}. However, due to their higher cohesive energies, transition metals tend to form cluster and their decoration over the monolayers would practically challenging \cite{song2015electric}.
Therefore, the clustering issue of transition metals considerably mitigates  the adsorption of H$_2$ molecules \cite{song2015electric,wang2014metal}. In addition, Alkali metals being lighter with respect to transition ones, would be preferred due to higher the hydrogen storage density \cite{sosa2021light}.

\begin{figure*}[t]
	\centering
	\includegraphics[width=0.9\linewidth]{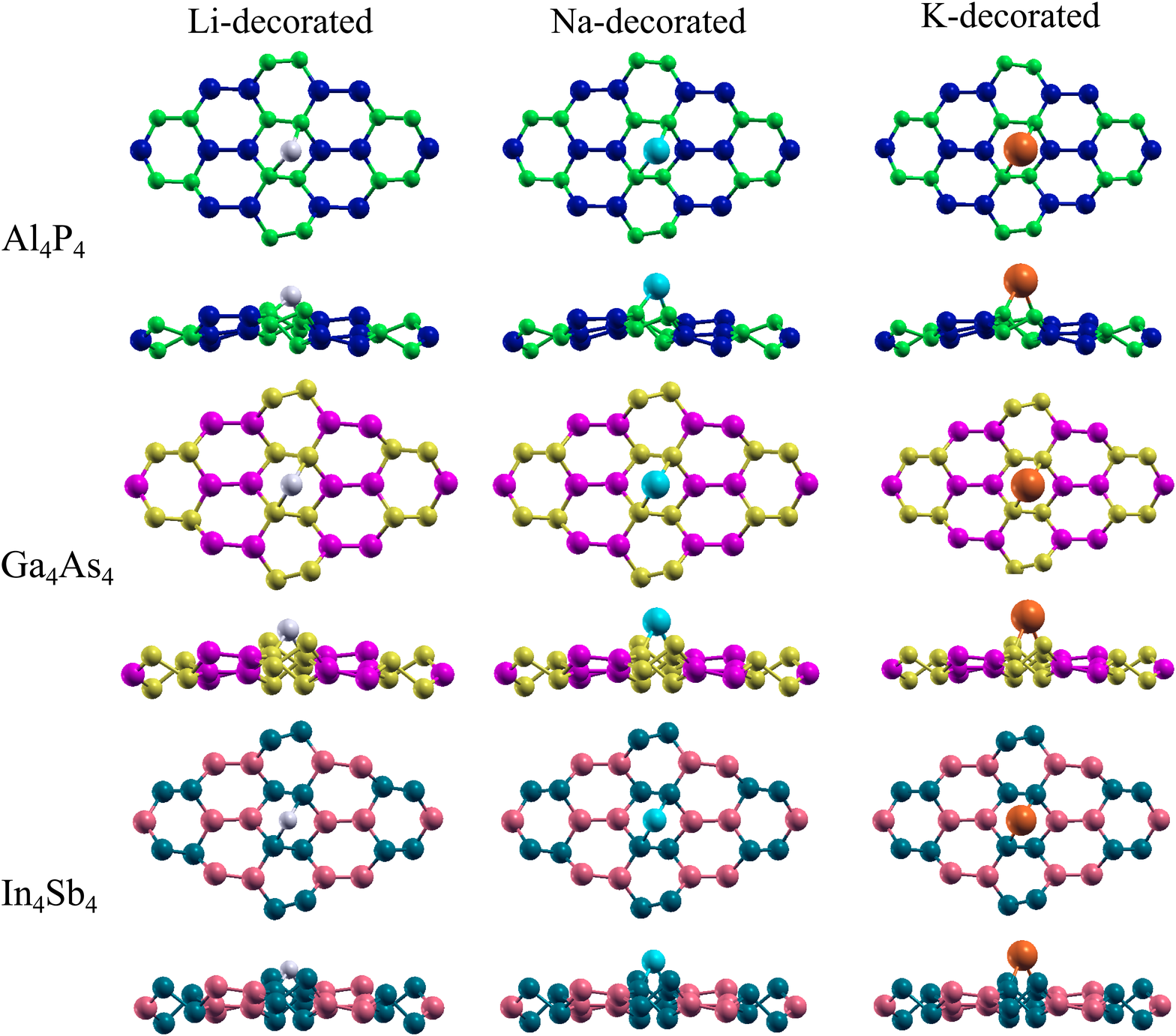}
	\caption{Top and side views of M-decorated ($\mathrm{M}=\mathrm{Li}$, Na and K) III$_4$-V$_4$ ($\mathrm{III}=\mathrm{Al}$, Ga and As and $\mathrm{V}=\mathrm{P}$, As and Sb). As it is evident, the location of alkaline atoms is at the center of the underlying hexagon. This atom is bridging over two opposite group-V atoms which are located above the plane of group-III atoms in the pristine structure.  It is obvious that the larger adatoms stand at a higher location with respect to the monolayer. }
	\label{fig:metal}
\end{figure*}


Previously. porous graphene has been functionalized using titanium atoms to absorb a maximum of eight
hydrogen molecules \cite{tan2018porous,sharifuzzaman2020smart,huo2021boron,yuan2018hydrogen}. In addition, the Pd- and Cu-decorated graphene together with NH and boron doping were also proposed to absorb the
hydrogen molecules \cite{choudhary2016first,singla2021enhanced,singla2021synergistic,singla2021effect}. 
Furthermore, Lithium, sodium and potassium elements were also deployed to
functionalize the 2D materials to enhance their interactions
with the hydrogen molecules \cite{sosa2021light,yadav2015first,khan2021computational}. In this regard, 
lithium, sodium and potassium atoms have been used for decoration of several different monolayers such as SiC, C$_2$N, WS$_2$, BNC and germanene to enhance the interaction with the hydrogen molecules \cite{hashmi2017ultra,qin2018high,song2015hydrogen,sarvazad2019hydrogen,sosa2021alkali,arellano2021hydrogen,yuan2021first,rahimi2020first,zhang2021reversible,huang2021li}. 
The Na-decorated borophene was also reported to absorb up to four hydrogen molecules per unit cell \cite{wang2018first,arellano2021hydrogen}. In another study, Na-decorated borophene is shown to be capable of absorbing seven hydrogen molecules per Na atom, although average hydrogen binding is reported instead of absolute value \cite{zhang2019hydrogen}.

\begin{figure*}[t]
	\centering
	\includegraphics[width=0.98\linewidth]{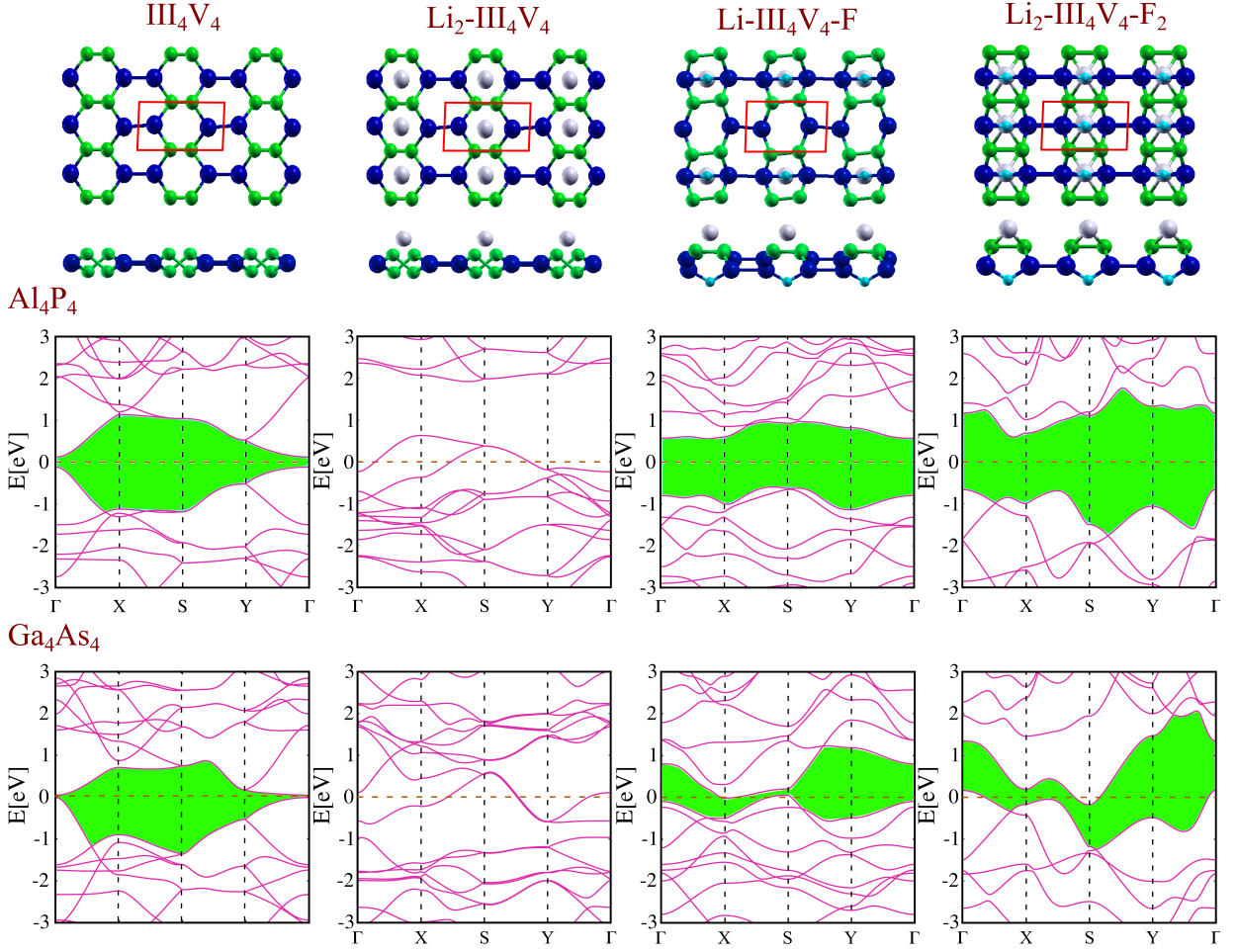}
	\caption{The structure and band structure of Al$_4$P$_4$ and Ga$_4$As$_4$ monolayers along with its lithium-decorated and fluorine-functionalized lithium-decorated.  }
	\label{fig:bands}
\end{figure*}



Previously, an external electric field has been deployed to enhance the hydrogen storage binding energy of Li-decorated graphene, Ca-decorated MoS$_2$ and alkaline metal decorated GaS \cite{zhang2016electric,song2015electric,mishra2020enhancement}. The role of electric field is to push the charge away from the decorating atoms. Therefore, with a lower charge density around the decorating atom, it would demonstrate higher binding to the hydrogen molecules. This mitigation of charge density over the adatoms can also be achieved using high electronegative elements. These elements should be deployed on the opposite side of the monolayer in order to absorb the charge density of the decorating atom.

Here, we demonstrate the hydrogen storage capability of alkali-decorated III$_4$-V$_4$ monolayers. In this regard, the maximum number of absorbed hydrogen molecules and their corresponding binding energies is reported. In order to enhance the hydrogen storage capability, the fluorine adatom is deployed to the opposite side of alkali-decorated III$_4$-V$_4$ monolayers.

\section{Computational details}

The density functional theory calculations which are embedded within the Quantum ESPRESSO bundle \cite{giannozzi2009quantum,giannozzi2017advanced} is applied here. Projector augmented plane waves (PAW) pseudopotentials are deployed as an exchange-correlation potential inside the generalized gradient approximation (GGA) parametrization of the PerdewBurke-Ernzerhof (PBE) \cite{perdew1996generalized}. The norm-conserving pseudopotentials defined the electron-ion interactions \cite{goedecker1996separable}. Furthermore, for the plane-wave base expansion, the energy cutoff of 50 Ry is applied in all calculations. A $12\times12$ Monkhorst-Pack k-point grid is employed in the measurement. Total geometry optimization is first performed until the pressures on the atoms become smaller than 0.001 eV/$\AA$ and the overall energy difference decays bellow $10^{-8}$ eV.

Applying the first principle theory, the transition path of the hydrogen absorption is studied using Gaussian 09 package. In addition to the geometry optimization, the binding energy calculations are also presented to reassure the structure stability. The B3LYP hybrid functional is deployed for the density functional calculation. This functional is extensively utilized in 2D structural materials, specifically for group V elements \cite{bhuvaneswari2019investigation,kokabi2022deep}. The basis of LanL2DZ is also picked for the calculations which has produced interesting results for group V elements commonly \cite{check2001addition,kokabi2020electronic}. The optimization iterations are followed until the maximum force constraints of $ 2\times 10^{-6}$ atomic unit and the highest displacement of $6\times 10^{-6}$ atomic unit. These criteria are assumed to be a very tight constraint. Force constants are computed in each optimization iteration. The symmetry is turned off during the calculations. An ultra-fine integration grid and exclusive quadratic convergence are assumed for the \emph{ab-initio} calculations.  The geometric structure, lowest-energy structure, stability and electrical characteristics of the neutral and singly charged nanosheets are also investigated.

\begin{figure*}
	\centering
	\includegraphics[width=1.0\linewidth]{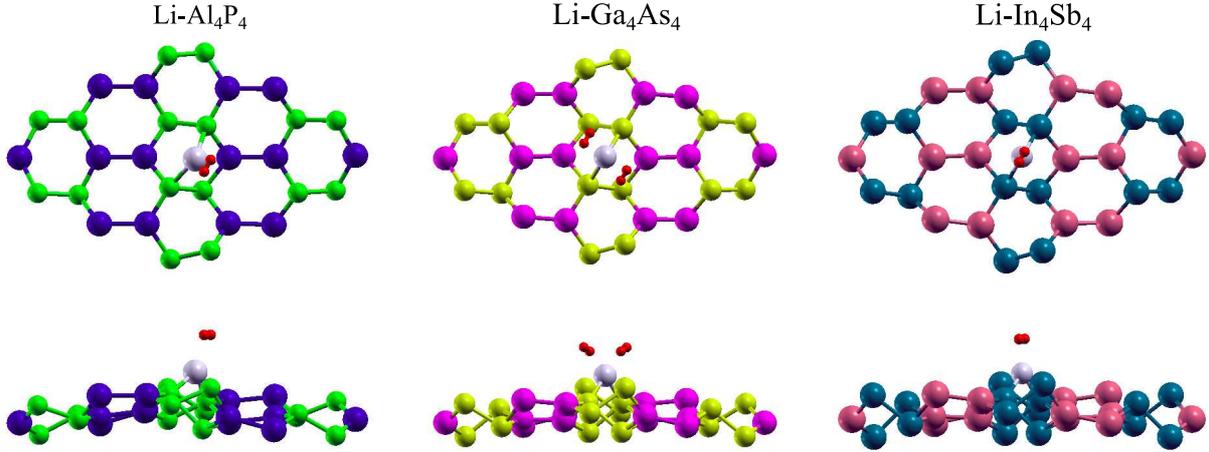}
	\caption{The schematic representation of hydrogen molecule absorbed to the Li-decorated III$_4$-V$_4$ ($\mathrm{III}=\mathrm{Al}$, Ga and As and $\mathrm{V}=\mathrm{P}$, As and Sb) from top and side views. }
	\label{fig:metal-h}
\end{figure*}

\section{Results and discussion}

Recently, a new binary monolayer compound of group-III and V elements with semi-buckled configuration and unique electronic properties has been proposed \cite{kokabi2021structural}. The decorated structures of III\textsubscript{4}-V\textsubscript{4} with three different alkaline metals of lithium, sodium and potassium are drawn in Figure \ref{fig:metal}. In these structures, the alkaline atom is bridging between two opposite group-V sites. These two group-V atoms are located above the plane of group-III atoms in the pristine structure. In addition, the location of alkaline atoms is at the center of the underlying hexagon. While the metal atoms bridge over two group-V atoms, these two atoms are located on the opposite sides of the hexagonal loop. Among all possible locations for the metallic adatom, the one in the middle of the hexagonal loop has the lowest possible energy. It should also be mentioned that the larger adatoms stand at a higher location with respect to the monolayer. No sensible distortion is observed for the underlying hexagonal loop after decoration.

In Table \ref{tab:tab1}, the binding energy of three different alkaline metals to three different III\textsubscript{4}-V\textsubscript{4} monolayer is presented. Maximum binding is observed for adding lithium atom to Ga$_4$As$_4$ compound. Fourth-row monolayer compound results in higher binding to alkaline elements. On the other hand, lithium is better attached to these monolayers while sodium demonstrates the least attachment. The trend from lithium to potassium can be explained by two contradicting factors of atomic radius and chemical activity. Heavier alkaline metals have higher chemical activity and are better attached to monolayer. On the other hand, the larger atomic radius makes them unfavorable for the monolayer due to lattice mismatch. 

%
%

In the previous works, it is shown that the electric field can enhance the hydrogen storage capability of alkaline decorated monolayers \cite{zhang2016electric,song2015electric,mishra2020enhancement}. This might be associated with the charge transfer from lithium to the underlying monolayer. This charge transfer effect can be achieved by a high electronegative element such as fluorine. Therefore, in order to enhance the hydrogen storage capability of Li-III\textsubscript{4}-V\textsubscript{4}, here we functionalize these monolayers with an additional fluorine atom on the opposite side of lithium element (M-III$_4$V$_4$-F). The binding energy of the fluorine atom to the Li-Al$_4$P$_4$ and Li-Ga$_4$As$_4$ is 2.312 and 2.019 eV, respectively.

In Figure \ref{fig:bands}, four different configurations and their corresponding band structures are depicted. The first structure is the pristine semi-buckled Al$_4$P$_4$. The second one is lithium-decorated of Al$_4$P$_4$ monolayer in which the lithium atoms are located in the middle of P-dominated hexagonal loops. The M-decoration together with F-functionalization can be accomplished by either all V-dominated hexagonal loops or just non-neighboring V-dominated ones. The unit cells corresponding to these two different decoration types are shown in Figure \ref{fig:bands}. In addition, this figure contains the band structures of these two structures. As it is obvious from the plotted band structures, both decoration types result in semiconducting behavior unlike metallic characteristic of Li-Al$_4$P$_4$. Here, the fluorine atom is placed opposite of lithium adatom on the other side of the monolayer. Fluorine is bridging between two group-III elements and pulling them selfwards. This further pushes all group-V elements M-wards. In contrast with M-decorated, here, the alkaline atom is bridging four group-V atoms. The side view of Li-Al$_4$P$_4$-F and Li$_2$-Al$_4$P$_4$-F$_2$ can be discriminated from the c-axis position of Al atoms. As it is obvious from the figure, all aluminum atoms of Li$_2$-Al$_4$P$_4$-F$_2$ structure are in-plane. In contrast, Li-Al$_4$P$_4$-F has two separate plane of aluminum atoms located in parallel to each other. In addition, the band structure of Ga$_4$As$_4$ is also depicted in this figure. The pristine Ga$_4$As$_4$ shows a small bandgap at \textGamma-point. Li-Ga$_4$As$_4$-F and Li$_2$-Ga$_4$As$_4$-F$_2$ demonstrate semi-metallic characteristic where the conduction band minimum (CBM) and valence band maximum (VBM) are located at two distinct points.

The corresponding band structures of two different M-III$_4$V-$_4$-F structures are depicted in Figure \ref{fig:bands}. The pristine Al$_4$P$_4$ demonstrates a small direct bandgap of 0.239 eV at gamma-point \cite{kokabi2021structural}. As we previously mentioned, no bandgap is observed for M-III$_4$-V$_4$ structure since no closed-shell is formed for valence electrons. On the other hand, the bandgap is expected for the closed-shell structures of Li-Al$_4$P$_4$-F and Li$_2$-Al$_4$P$_4$-F$_2$. From Figure \ref{fig:bands}, an indirect bandgap of 1.25 eV is obvious for Li-Al$_4$P$_4$-F. Furthermore, Li$_2$-Al$_4$P$_4$-F$_2$ structure demonstrates an indirect bandgap of 1.23 eV. The two bandgap values are much higher than that of pristine structure. Therefore, simultaneous decoration by lithium and fluorine elements would significantly enhance all Al$_4$P$_4$ bandgap. 

\begin{figure}
	\centering
	\includegraphics[width=1.0\linewidth]{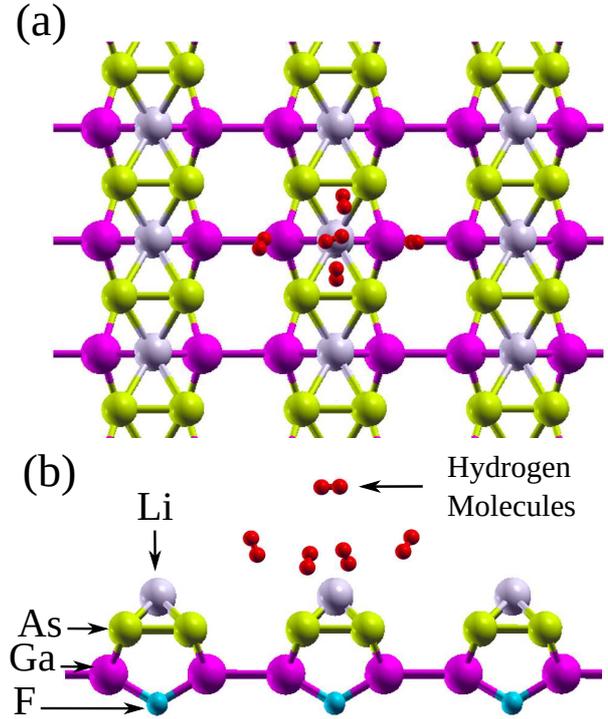}
	\caption{(a) Top and (b) side view schematic of five hydrogen molecules absorbed to the F-functionalized Li-Ga$_4$-As$_4$.}
	\label{fig:gaas-5h2}
\end{figure}

\begin{table}[]
	\centering
	\caption{Alkaline-metal binding energy to the underlying monolayer together with the binding energy of hydrogen molecule to the Li-decorated monolayers. \label{tab:tab1}}
	\begin{tabular}{lllll}
		\hline
		& Li       & Na       & K        & Li+H2    \\
		\hline
		\hline
		Al$_4$P$_4$                    & 1.833 eV &  1.402 eV& 1.573 eV & 0.007 eV \\
		Ga$_4$As$_4$                   & 2.008 eV & 1.584 eV & 1.634 eV & 0.114 eV \\
		In$_4$Sb$_4$                   & 1.744 eV & 1.344 eV & 1.502 eV & 0.111 eV \\
		\hline
	\end{tabular}
\end{table}

\begin{table*}[]
	\centering
	\caption{The hydrogen molecule binding energy to the Li-decorated Ga$_4$As$_4$.\label{tab:tab2}}
	\begin{tabular}{llllll}
		\hline
		& 1        & 2        & 3         & 4        & 5        \\
		\hline
		\hline
		Li-ave                   & 0.114 eV & 0.098 eV & 0.019 eV  &          &          \\
		Li-abs                   & 0.114 eV & 0.082 eV & -0.138 eV &          &          \\
		Li-F-ave                 & 0.272 eV & 0.175 eV & 0.125 eV  & 0.1 eV   & 0.083 eV\\
		Li-F-abs                 & 0.272 eV & 0.078 eV & 0.026 eV  & 0.022 eV & 0.016 eV \\
		\hline
	\end{tabular}
\end{table*}

Figure \ref{fig:metal-h} shows the schematic of binding hydrogen molecules to M-decorated III\textsubscript{4}-V\textsubscript{4} monolayer. H-H bonds in the hydrogen molecules are aligned with the more distant III-V bonds. No sensible deformation is observed for the underlying monolayer after adding hydrogen molecules. The average molecular distance between the hydrogen molecules and lithium atom is 2.25 \text{\normalfont\AA}. The first binding energy to lithium atom is listed in Table \ref{tab:tab1}. Li-Al$_4$P$_4$ demonstrates the least binding energy to the hydrogen molecule. Li-Ga$_4$As$_4$ and Li-In$_4$Sb$_4$ show almost similar attachment to the hydrogen molecule while Li-Ga$_4$As$_4$ is slightly better.

In Figure \ref{fig:gaas-5h2}, Li-Ga\textsubscript{4}As\textsubscript{4}-F structure storing four and five hydrogen molecules are depicted. In the case of storing four hydrogen molecules, these molecules form a rhombic structure around the alkaline metal. Two of these molecules are closer to the alkaline atom while the two others are more distant. In the case of Li-Ga$_4$As$_4$-F, the shorter and longer distances are 2.26 and 3.82 \text{\normalfont\AA}, respectively. When Li-Ga\textsubscript{4}As\textsubscript{4}-F structure is storing five hydrogen molecules, the previous rhombic structure is still evident while the fifth molecule is added to the top of the mentioned structure exactly over the alkaline atom. In the case of Li-Ga$_4$As$_4$-F, the shorter and longer distances are 2.26 and 4.32 \text{\normalfont\AA}, respectively and the top molecule is located 4.51 \text{\normalfont\AA} apart from the lithium atom. 

The average binding energy can be defined as
\begin{equation}
E_{ave}=\frac{E_{\mathrm{Monolayer}}+nE_{\mathrm{H_2}}-E_{\mathrm{nH_{2}/Monolayer}}}{n},
\end{equation}
where $E_{\mathrm{Monolayer}}$ and $E_{\mathrm{H_2}}$ denote the total energy of the monolayer and hydrogen molecule, respectively.   $E_{\mathrm{nH_{2}/Monolayer}}$ is the total energy of the monolayer when adsorbed n hydrogen molecules. Furthermore, for better insight, we also report the absolute binding energy of $n$th hydrogen molecule which is
\begin{equation}
E_{abs}=E_{\mathrm{(n-1)H_{2}/Monolayer}}+E_{\mathrm{H_2}}-E_{\mathrm{nH_{2}/Monolayer}}.
\end{equation}

The variation of the binding energy from the first to fifth hydrogen molecules for Li-Ga$_4$As$_4$-F is presented in Table \ref{tab:tab2}. Li-Ga$_4$As$_4$ can store up to two hydrogen molecules with the maximum binding of 0.114 eV. On the other hand, Li-Ga$_4$As$_4$-F is capable of storing up to four hydrogen molecules with the maximum binding of 0.272 eV. Therefore, the added fluorine atom enhances both the binding energy and maximum number of stored hydrogen molecules. Interestingly, the first binding energy of Li-Ga$_4$As$_4$-F is more than twice that of Li-Ga$_4$As$_4$.

\section{Conclusion} 
Here, the effects of fluorine functionalization on the hydrogen storage performance of alkaline decorated III\textsubscript{4}-V\textsubscript{4} monolayers are studied. Among the considered structures, the maximum binding is obtained for adding the lithium atoms to Ga$_4$As$_4$ compound. The M-decoration is achieved by either all V-dominated hexagonal loops or just non-neighboring V-dominated ones. A maximum of two hydrogen molecules can attach to the M-decorated III\textsubscript{4}-V\textsubscript{4} monolayer.

In order to have charge transfer effect and consequently enhancement of hydrogen storage capability, a high electronegative fluorine atom is deployed for the functionalization of M-III\textsubscript{4}-V\textsubscript{4}. Therefore, we functionalize these monolayers with additional fluorine atom on the opposite side of lithium element and fluorine is bridging between two group-III elements. While M-decorated III\textsubscript{4}-V\textsubscript{4} demonstrates metallic behavior, functionalizing these monolayers with fluorine results in bandgap opening and semiconducting characteristics. The bandgap for Li-Al$_4$P$_4$-F is 1.23 eV which is very close to the solar peak. In addition, Fluorine functionalizing Li-III\textsubscript{4}-V\textsubscript{4} structure can enhance the hydrogen molecule storage capability from two to four. The binding of the hydrogen molecule to the alkaline atom is also improved from 0.114 to 0.272 eV.  



	
\end{document}